\documentclass[preprint,
preprintnumbers,amsmath,amssymb,superscriptaddress]{revtex4-2}
\usepackage{fullpage,amsthm,amsmath,amsfonts,bm,times,color,bbm}
\usepackage{graphicx,subfigure}
\usepackage{float}
\usepackage[english]{babel}
\usepackage{mathtools}
\def\degree{\hbox{$^\circ$}}

\def\eqa{\begin{eqnarray}}
\def\eea{\end{eqnarray}}
\newcommand{\eq}{\begin{equation}}
\newcommand{\ee}{\end{equation}}

\makeatother
\def\degree{\hbox{$^\circ$}}
\newcommand{\el}{El\ Ni\~{n}o}
\newcommand{\lanina}{La\ Ni\~{n}a}
\makeatletter

\begin{document}

\title{Teleconnections among Tipping Elements in the Earth System}

\author{Teng Liu}
%\email{}
\affiliation{School of Systems Science, Beijing Normal University, Beijing 100875, China}
\author{Dean Chen}
%\email{}
\affiliation{School of Systems Science, Beijing Normal University, Beijing 100875, China}
\affiliation{Institute for Atmospheric and Earth System Research/ Physics, Faculty of Science,
University of Helsinki, 00014 Helsinki, Finland}

\author{Lan Yang}
%\email{}
\affiliation{School of Systems Science, Beijing Normal University, Beijing 100875, China}

\author{Jun Meng}
%\email{}
\affiliation{School of Science, Beijing University of Posts and Telecommunications, Beijing 100876, China}
\affiliation{Potsdam Institute for Climate Impact Research, Potsdam 14412, Germany}

\author{Zanchenling Wang}
%\email{}
\affiliation{Yuanpei College, Peking University, Beijing 100871, China}

\author{Josef Ludescher}
%\email{}
\affiliation{Potsdam Institute for Climate Impact Research, Potsdam 14412, Germany}

\author{Jingfang Fan}
\email{jingfang@bnu.edu.cn}
\affiliation{School of Systems Science, Beijing Normal University, Beijing 100875, China}
\affiliation{Potsdam Institute for Climate Impact Research, Potsdam 14412, Germany}

\author{Saini Yang}
%\email{}
\affiliation{State Key Laboratory of Earth Surface Processes and Resource Ecology, Beijing Normal University, Beijing 100875, China}
\affiliation{School of National Safety and Emergency Management, Beijing Normal University, Beijing 100875, China}

\author{Deliang Chen}
%\email{}
\affiliation{Department of Earth Sciences, University of Gothenburg, Gothenburg 40530, Sweden}

\author{J\"urgen Kurths}
%\email{Juergen.Kurths@pik-potsdam.de}
\affiliation{Potsdam Institute for Climate Impact Research, Potsdam 14412, Germany}
\affiliation{Department of Physics, Humboldt University, 10099 Berlin, Germany}

\author{Xiaosong Chen}
%\email{}
\affiliation{School of Systems Science, Beijing Normal University, Beijing 100875, China}

\author{Shlomo Havlin}
%\ead{havlin@ophir.ph.biu.ac.il}
\affiliation{Department of Physics, Bar Ilan University, Ramat Gan 52900, Israel}
\author{Hans Joachim Schellnhuber}
%\ead{john@pik-potsdam.de}
\affiliation{Potsdam Institute for Climate Impact Research, Potsdam 14412, Germany}

\begin{abstract}
Tipping elements of the Earth system may shift abruptly and irreversibly from one state to another at tipping points, resulting in a growing threat to our society. Yet, it is not fully clear how to assess and quantify the influence of a tipping element and how to explore the teleconnections between different tipping elements. To fill this knowledge gap, we propose a climate network approach to quantitatively analyze the global impacts of a prominent tipping element, the Amazon Rainforest Area (ARA). We find that regions, such as, the Tibetan Plateau (TP) and West Antarctic ice sheet, are characterized by higher network weighted links and exhibit strong correlations with the ARA. We then identify a teleconnection propagation path between the ARA and the TP. This path is robust under climate change as simulated by various climate models of CMIP5 and CMIP6. In addition, we detect early warning signals for critical transition in the snow cover extent on the Tibetan Plateau by applying critical slowing down indicators, lag-1 autocorrelation and detrended fluctuation analysis. We find that the snow cover of the TP has been losing stability since 2008, revealing that the TP is operating like a tipping element and approaching a potential tipping point. We further uncover that various climate extremes between the ARA and the TP are significantly synchronized under climate change.
Our framework provides new insights into how tipping elements are linked to each other and into the potential
predictability of cascading tipping dynamics.
\end{abstract}
\date{\today}

%\flushbottom
\maketitle
\section{Introduction}
As a complex adaptive system, the Earth system has multiple potential tipping elements that may approach or exceed a tipping point in response to a tiny perturbation~\cite{lenton_tipping_2008}. Recently, Lenton~\textit{et al.} highlighted nine climate tipping points~\cite{lenton_climate_2019}, such as the collapse of the Greenland ice sheet, the loss of the Amazon   rainforest~\cite{lovejoy2019amazon,taubert_global_2018}, the slowdown of the Atlantic Meridional Overturning Circulation~\cite{boers_observation-based_2021}, the instability of the West Antarctic ice sheet~\cite{garbe_hysteresis_2020},  the decay of the Greenland ice sheet~\cite{meredith2019polar} that have been activated in the past decade, and therefore urgent political and economic action is needed to reduce greenhouse gas emissions in order to prevent key tipping elements from tipping. Anthropogenic forcing is considered as one of the main factors for pushing several large-scale ‘tipping elements’ to exceed their tipping points~\cite{ghil_physics_2020}, which may cause abrupt and irreversible destabilizing effects on the Earth system~\cite{lenton_early_2011}, such as cryosphere changes, sea level rise, heatwave-drought and so on~\cite{IPCC6,Zhang2020}. Interactions between the different tipping elements may either have stabilizing or destabilizing effects on the other subsystems, potentially leading to cascades of abrupt transitions~\cite{scheffer_anticipating_2012}. Following the rising awareness of a highly interconnected world, tipping cascades as possible links between tipping elements are increasingly discussed~\cite{Klose2021}. Using palaeoenvironmental records, Brovkin~\textit{et al.} illustrated how abrupt changes cascaded through the Earth system in the past 30 kyr~\cite{Brovkin2021}. Steffen~\textit{et al.} pointed out that tipping cascades could be formed when the global temperature reaches a threshold, affecting the trajectories of the Earth system in the Anthropocene~\cite{Steffen2018}. However, a quantitative and systematic analysis framework on how the Earth system can be influenced by tipping elements is still lacking, especially for identifying connections among tipping elements. While preliminary studies conceptually proposed possible connections between tipping elements~\cite{lenton_climate_2019,martin2021}, how these tipping elements are influenced by the mode of others and what are the teleconnection paths are still open questions.\\

\noindent Here we focus on the prominent and well-known tipping element--the Amazon rainforest area. Tropical forests play a vital role in the global carbon cycle, which account for one-third of the gross primary production and host more than half of the known worldwide species. It has been reported that human activities and climate change have perturbed the stability of the tropical forest-climate equilibrium, resulting in extreme losses of tropical forests and biodiversity~\cite{gibson_primary_2011}. The annual deforestation rate of tropical forests has been 0.5\% since the 1990s~\cite{achard_determination_2014}, accounting for 32\% of the global tree loss~\cite{hansen_high-resolution_2013}. A recent empirical study suggests that the observed tropical forest fragmentations, including in the Americas, Africa and Asia–Australia, are approaching their tipping points~\cite{taubert_global_2018}. Particularly striking is the deforestation in the Amazon--the world’s largest rainforest--harbors which consists of nearly a quarter of the world’s terrestrial species. Southeast Amazonia has even become a net source of carbon emission during the dry season because of the deforestation and climate change~\cite{gatti_amazonia_2021}. Boulton~\textit{et al.} pointed out that more than three-quarters of the Amazon rainforest has been losing resilience since the early 2000s, consistent with the approach to a critical transition~\cite{boulton_pronounced_2022}. Extreme precipitation and prolonged dry seasons have become much more frequent~\cite{nobre_land-use_2016}, which have been regarded as warnings of climate tipping points~\cite{lovejoy2019amazon}. The Sixth Assessment Report of IPCC WGI highlighted that continued deforestation and a warming raise the probability that the Amazon will cross a tipping point into a dry state~\cite{IPCC6}. However, global influences of rainforest dieback in the Amazon are still little known. Here, we develop a climate network-based framework to systematically study the global impacts of the Amazon area.\\

\noindent Networks have proven to be a versatile tool to explore the dynamical and structural properties of complex systems in many disciplines, including in physical, biological, ecological and social sciences~\cite{newman2010networks}.  The application of network theory on the complex climate system led to the birth of the idea of~\textit{climate networks} (CN) in which geographical locations on a longitude-latitude grid become network nodes, while the degree of similarity, or `connectedness', between the climate records obtained at two different locations determines whether there is a network link between these locations~\cite{tsonis_architecture_2004}. The climate-network framework has been applied successfully to analyse, model, and predict various climate phenomena, such as ENSO~\cite{ludescher_improved_2013}, extreme rainfall~\cite{boers_complex_2019}, Indian summer monsoon~\cite{NetworkMonsoon}, Atlantic meridional overturning circulation~\cite{mheen_interaction_2013}, Atlantic multidecadal oscillation~\cite{feng_are_2014} and others~\cite{fan_statistical_2021}. The power of the CN lies in its ability to map out the topological features and pattern that are related to the physics of the dynamical climate variability.\\

\noindent In the present study, we construct a series of dynamical and physical climate networks based on the global near-surface air temperature field. The directed links from the ARA (see Fig.1 location labeled 6) to regions outside the ARA are defined as `in'-links.  The in-weighted climate network enables us to obtain a map of the global impacts of ARA, in particular, to study the impacts in specific regions, such as other tipping elements. In particular, we uncover that there exists a robust negative teleconnection between the ARA and the Tibetan Plateau (TP) (see Fig.1 location labeled 10), known as the third pole of the Earth~\cite{yao_recent_2019}. We further explore the potential propagation pathway of the ARA-TP teleconnection. This work provides a novel concise and systematic framework to investigate the potential teleconnection among the tipping elements, and can potentially be used to predict the abrupt changes caused by the tipping cascading in the Earth systems.\\

\section{Results}

\subsection{Strongly localized global impact pattern of the ARA}

To reveal the global impact of the ARA, we divide the nodes of climate networks into two subsets. One subset includes the nodes within the ARA (1374 nodes,  $12^o N-35^o S$, $30^o-90^o W$ land area) and the other contains the nodes outside the ARA (63786 nodes). To quantify the overall impact of the ARA, for a given outside node $j$, we define its in-weight ($IN(C_j)$) and in-strength ($IN(W_j)$) as the sum of the weights and strengths of all in-links (see Methods). A larger (smaller) positive (negative) value of $IN(C_j)$ and $IN(W_j)$ indicate stronger (weaker) warming (cooling) due to the influence of the ARA; likewise, out-weights ($OUT(C_i)$ and $OUT(W_j)$) are introduced to quantify the overall impacts to the ARA. We present the influence pattern between the ARA and the outside area in Fig. S1a, b for the past 40 years (1979-2018). We find that some regions such as the Mid-Atlantic, the Arctic, and the Indian Ocean, either by warming or cooling, are identified by relatively higher in-weights (Fig. S1a). It has been reported that the climate variability in the ARA is significantly affected by the phase of ENSO~\cite{cai_climate_2020}, this influence is also confirmed by the high intensity of the tropical Pacific region in out-weight distributions (Fig. S1b). In particular, we find that the Ni\~{n}o 3.4 region shows a strongly positive influence on the ARA.\\

\noindent There exist strong connections between ARA and ENSO. To further address whether the influence pattern of ARA varies between ENSO periods and normal years, we analyze and compare the global distributions of the total in-degrees, $IN(N)$; in-weights, $IN(C)$; in-strengths $IN(W)$, among one typical \el~year (1997), one typical \lanina~year (1998) and one normal year (1996). The result is shown in Fig. 2. We find that during \el~ and \lanina~events, the overall global area that is influenced by the ARA becomes smaller, whereas the impacts in these more limited ranges become stronger. This enhanced impact in localized regions is demonstrated in Fig. 2a-f, which compares the global distributions of the normal year in Fig. 2g-i. This localized influence pattern during the ENSO period is supported by more examples, as shown in Fig. S2.

\subsection{Negative teleconnection between ARA and TP}

The regions of localized activity vary from one year to another. Next, we study the variability of the regions that are influenced by ARA. For each year, we consider the significantly influenced nodes, i.e., with the top 10\% of the in-degree and negative in-weight. We then define the frequency of the nodes appearing during the past 40 years from 1979 to 2018 as, $F(N)$ and $F(C)$ (see Methods for details). The results are shown in Fig. 3a, b for $F(N)$ and $F(C)$, respectively. Higher values indicate stronger and more persistent impacts of ARA. Remarkably, we find that the intensity of the nodes within the TP region is high for both $F(N)$ and $F(C)$, and the spatial pattern of these nodes agrees well with the cartographic boundary (the dashed yellow lines in the right panels of Fig. 3a, b) and shape of the TP. A similar pattern can also be obtained from the $F(W)$ (Fig. S3), which serves as a cross-check of the result of $F(C)$. A typical cross-correlation function of two nodes, one in ARA and the other in TP, is presented in Fig. S4, which indicates a significant negative teleconnection. Besides the TP, we also observe that there exist strong negative connections between the ARA and the West Antarctic, which is known as a tipping element~\cite{lenton_climate_2019}. It indicates that our method can be used to reveal the teleconnection between specific tipping elements.\\

\noindent To demonstrate that these results are not accidental, we analyzed randomized versions of $F(N)$ and $F(C)$, which we obtained by reshuffling the temperature records at each site 100 times. This way, we destroy the correlations between different nodes (one example of the cross-correlation results of this NULL model can be found in Fig. S5c, d). Our results shown in Fig. 3c, d indicate that the values of $F(N)$ and $F(C)$  for the nodes in the TP region have a 95\% confidence level compared with the Null-model.

\subsection{Robust propagation pathway of the teleconnection between the ARA and TP}

Teleconnections describe remote connections between components of the complex climate system and reflect the transportation of energy or materials on global scale~\cite{Zhengyu2007}. In the following, we will identify the actual path of the teleconnection between the ARA and the TP by using the climate network analysis. We choose 726 latitude-longitude grid points as CN nodes~\cite{NetworkMonsoon}, such that the globe is covered approximately homogeneously. We follow Zhou \textit{et al.}~\cite{zhou_teleconnection_2015} and detect a minimal total cost function of the direct links (see Methods for details) from the ARA to the TP. We show a potential propagation path for this teleconnection in Fig. 4a, and find out that it can be roughly divided into three parts. The first part is from the center of South America to the south of Africa, the second one is from the south of Africa to the Middle East, and the last part is from the Middle East to the TP. From a meteorological perspective, the existence of such a teleconnection between the ARA and TP is no coincidence, and can be well explained by the main atmospheric and oceanic circulation.\\

\noindent On the one side, the orography of the eastern coast of the South American continent is prone to the formation of an anticyclone at mid-latitude due to the interaction with mid-latitude westerlies~\cite{leduc1984connaitre}. The anticyclonic circulation produces and brings warm winds from the east coast of South America to the South of Africa (around 30 \degree S).  On the other side, there is an intertropical convergence zone that controls the African monsoon driving the wind from south to north of Africa in this regime~\cite{nicholson_itcz_2018}. Finally, the physical mechanism of the path from the Middle East to the TP may be linked to the northern hemispheric middle latitude Westerlies~\cite{kong_interaction_2020}.\\

\noindent To examine the data dependence of our approach, we perform the same CN-based analysis to determine the optimal pathway of teleconnection~\cite{zhou_teleconnection_2015} between the ARA and TP by using a different reanalysis datasets, i.e., the ERA5 with 2m surface temperature as well as the NCEP/NCAR reanalysis with 1000hPa, 2m surface temperature dataset. All results are presented in Fig. S6, and suggest that the propagation pathway of the teleconnection between the ARA and TP is quite robust and independent of datasets. Furthermore, different starting and ending nodes within ARA and TP have been selected by using the same analysis, and we find that the optimal pathway between ARA and TP is still very stable (Fig. S8). This further reveals that our proposed physical mechanism of this teleconnection path is convincing and effective.\\

\noindent Anthropogenic climate change  has  led  to a widespread  shrinking  of  the  cryosphere, rising  global mean-sea levels, an increasing number of tropical cyclones,  and associated cascading impacts~\cite{meredith2019polar}. A critical question, then, is how climate change could affect the nature of the path of this teleconnection? We thus investigate the response of the teleconnection path to global warming by using the Coupled Model Intercomparison Project phase 5 (CMIP5) and phase 6 (CMIP6) models under the ECP8.5 (approximately equivalent to the shared socioeconomic pathway 5-8.5) emission scenario from 2006 (CMIP5) and 2016 (CMIP6) to 2100. We chose 15 CMIP5 models and 15 CMIP6 models and summarize the details in TABLE S1 and S2. To identify the response of the teleconnection pathway under the global warming condition, for simplicity but without loss of generality, we compare, in Fig. 4b-e, the  path from ARA to TP for the first and last 40 years of the 21st century, i.e., 2016-2056 (2006-2046 for CMIP5 datasets) vs. 2060-2100.  Interestingly, we observe that the overall pattern of this teleconnection pathway from the ARA to TP is quite stable across most models. Additionally, we calculate the propagation time by summing the time lags for each step, and find that the time delay is around 15 days for most models (see Fig. S9).

\subsection{Propagation pathway of the teleconnection between the ARA and West Antarctic}

The accelerating and persistent mass loss in the Antarctic Ice Sheet, especially the West Antarctic Ice Sheet (WAIS), bring huge uncertainties in predicting and projecting a future sea-level rise in this and upcoming centuries. Model studies and palaeo-evidence~\cite{Feldmann14191} support that the widespread collapse of WAIS is a `low likelihood high impact' event, and some regions of the ice sheet may reach tipping points with a warming climate~\cite{pattyn_uncertain_2020}. In this study, the high intensity of the nodes within WAIS in Fig. 5b indicates the existence of stable negative teleconnection between these two well-known tipping elements, ARA and WAIS. The potential propagation paths for this teleconnection with different datasets have been shown in Fig. S10. We find that the path can be explained by the steady and strong ocean currents and westerly winds near the West Antarctic area. This implies that the transportation of substances, like dust and carbonaceous aerosols, is likely to cause this teleconnection. This hypothesis is supported by Antarctic pollution studies~\cite{McConnell2007} where the increased dust mobility in ARA caused by overgrazing and deforestation has a certain impact on West Antarctic pollution.

\subsection{Synchronization of extreme climate events between ARA and TP}
Since our results indicate that the teleconnection path is not affected by climate change (see Fig. 4), one key issue is how does the climate variabilities in between the ARA and TP synchronize in the presence of global warming? In the following, we focus on the synchronization of different extreme climate events between ARA and TP. We first analyze the global change by the fraction of days with above average temperature (Txgt50p) in every two decades (2021-2040, 2041-2060, 2061-2080 and 2081-2100) under four SSPs (Shared Socioeconomic Pathways), and notice a clear spatial synchronization of this indicator in the Amazon and the TP for all SSPs. Both the Amazon and the TP are the most sensitive areas globally in terms of the increase of Txgt50p, as shown in Fig. S12. Multi-model ensemble (MME) were applied here, based on six models from Global Producing Centers (GPCs) for long-range forecasts (LRFs) designated by the World Meteorological Organization (WMO), which has been confirmed to have higher simulation capacity than that of each model~\cite{kim_assessment_2021}. To quantify the spatial synchronization of these two regions, we performed Pearson’s correlation analyses on the Txgt50p values. Surprisingly, we find that the Pearson's $r$ reaches $~$0.92 for the Txgt50p in the Amazon and the TP, illustrating a significant positive correlation between these two regions. Similarly, we tested the other major temperature-related indicators, TMge10 (number of days with daily mean temperature equal to or above 10\degree C), and TNn (the monthly minimum of Tmin). Strong positive correlation exists between the TP and Amazon region for both TMge10 ($r=0.70$) and TNn ($r=0.74$). Meanwhile, a strong or moderate negative correlation exists in precipitation-related indicators, such as Prcptot (-0.54, Fig. S13), R20mm (-0.50, Fig. S14) and Rx5day (-0.20, Fig. S15). These three indicators represent annual total precipitation, number of days with precipitation above 20 mm and the maximum 5-day consecutive precipitation, respectively. These results support the finding that there exists some kind of physical connection between these two far-reaching regions. It is noticed that all precipitation-related indicators show a negative correlation, while temperature-related indicators show a positive correlation between the TP and Amazon.

\subsection{The TP is operating close to a tipping point}

The TP has attracted much attention due to its unique geological structure, irreplaceable role in global water storage and impact on the global climate system. The TP and its surroundings hold the largest amount of glaciers outside the Arctic and Antarctic, and the largest extent of high-altitude permafrost in the world. Thus, the TP is sensitive to climate change~\cite{you_warming_2021}. Numerous literatures have shown that the warming trend in recent decades at the TP is several times faster than the global average, and is similar to the trend of the arctic region~\cite{yao_recent_2019}. Projections have shown that this amplification of warming will continue under global warming, thereby increasing the occurrence of climate extremes~\cite{you_tibetan_2020}. Snow cover is a comprehensive indicator of the mean conditions of temperature and precipitation for an area. Especially for the TP, the snow cover can persist during all seasons over the high elevation area, and serve as a vital water source for the surrounding countries. Previous studies have reported that the snow cover can serve as a robust indicator to reveal the connection between the TP and other climate systems, including general circulation~\cite{ye1998role} and monsoon systems~\cite{YANG1996} over eastern and southern Asia. In particular, snow cover variability is an integrated indicator of climate change~\cite{Qin2006}, and thus can be a sensitive parameter reflecting the state change of the TP under global warming. In the following, we will detect the Early Warning Signals (EWS) based on the snow cover, and reveal that the TP has been losing stability and approaching a tipping point since 2008.\\

\noindent The `critical slowing down' (CSD) phenomenon has been suggested as one of the most important indicators of whether a dynamical system is getting close to a critical threshold~\cite{lenton_early_2011,ditlevsen_tipping_2010}. As the currently occupied equilibrium state of a system becomes less stable, the intrinsic rates at which a system recovers from small perturbations will become slower, and therefore the state of the system at any given moment becomes more like its past state. This loss of stability (defined as the return rate from perturbation) can be detected by increases in the lag-1 autocorrelation (AR(1)) coefficients and detrended fluctuation analysis (DFA) exponents~\cite{Peng1995}. These methods have been applied to quantify the CSD and anticipate tipping points~\cite{Livina2007,Dakos2008}. Here we focus on the temporal evolution of AR(1) coefficients and DFA exponents based on the time series of the snow cover fraction (SCF) over the whole TP area during 1990-2020 (see Methods). The Seasonal and Trend decomposition based on Loess (STL) method has been used to obtain the detrended time series and avoid false alarms (see Methods and Fig. 5a). By shifting a sliding 10-year window and calculating the AR(1) coefficients and DFA exponents for every step, we can retrieve the temporal changes of the AR(1) coefficients (Fig. 5b) and DFA exponents (Fig. 5c) in the past 31 years (1990-2020). The time series of the AR(1) coefficient shows a substantial increase over time, particularly since 2008 (the red line in Fig 5b). The DFA exponent has also been increasing since 2008 and is consistent with the AR(1) coefficient, as shown in Fig. 5c. We quantify the increasing tendency of AR(1) and DFA exponents by the Kendall rank correlation coefficient $\tau$ (see Methods), with 0.95 and 0.86, respectively. Both results of the Kendall $\tau$ support an obvious increase for AR(1) and DFA, which implies that the snow cover in TP has been approaching a tipping point since 2008.

\section{Discussion}
The persistent warming fueled by anthropogenic greenhouse gas emissions could push parts of the Earth system--tipping elements--into abrupt or irreversible changes, from collapsing ice sheets and thawing permafrost, to shifting monsoons and forest dieback~\cite{lenton_climate_2019}.  These climatic changes thus influence the nature of societies and the performance of economies~\cite{carleton_social_2016}. Most importantly, possible connections and cascading dynamics between different tipping elements have been proposed.  However, method to quantify the connections or teleconnections among possible tipping elements was lacking. To fill the gap, here, we have developed a network-based framework to reveal the global impact of a widely pronounced tipping element-- the Amazon Rainforest Area (ARA). We found that there is a global pattern of strongly localized impacts of the ARA, and some specific regions, such as the TP and West Antarctic, are strongly and persistently influenced by the ARA.  A robust teleconnection propagation path has been identified between the Amazon and TP. We proposed a possible physical mechanism underlying this path and associated it with the combination of: (1) the South Atlantic High, (2) the intertropical convergence zone and (3) the northern hemispheric middle latitude westerlies. The high degree of synchronization of the extreme events in the ARA and TP supports the existence of this teleconnection. Moreover, we provided strong support that the snow cover in TP (CDS phenomenon) has been losing stability, and is operating close to a tipping point. We thus provided evidence that the TP, which was previously overlooked, should play an extremely important role as a component in the exhaustive list of tipping elements~\cite{lenton_tipping_2008}.\\

\noindent Interactions and teleconnections between the different tipping elements may either have stabilizing or destabilizing effects on the other subsystems, potentially leading to cascades of abrupt transitions. In particular, in the context of climate change, disaster phenomena such as floods, droughts, wildfires, heatwaves, and sea level rises have become more frequent and threatening. Thus, global, national and regional preparedness and response to extreme weather are facing challenges. Climate adaptation failure has gained the greatest concern globally. Especially, the systemic risk induced by the interdependency among systems and cascading of adverse impact is the emerging key for climate adaptation. Our framework based on network theory provides a potential path to understand the linkage of tipping elements of the complex Earth system, which is particularly important for a systemic risk-informed global governance, and to improve our understanding of tipping points.

\clearpage
\begin{figure}
\centering
\includegraphics[width=1\textwidth ]{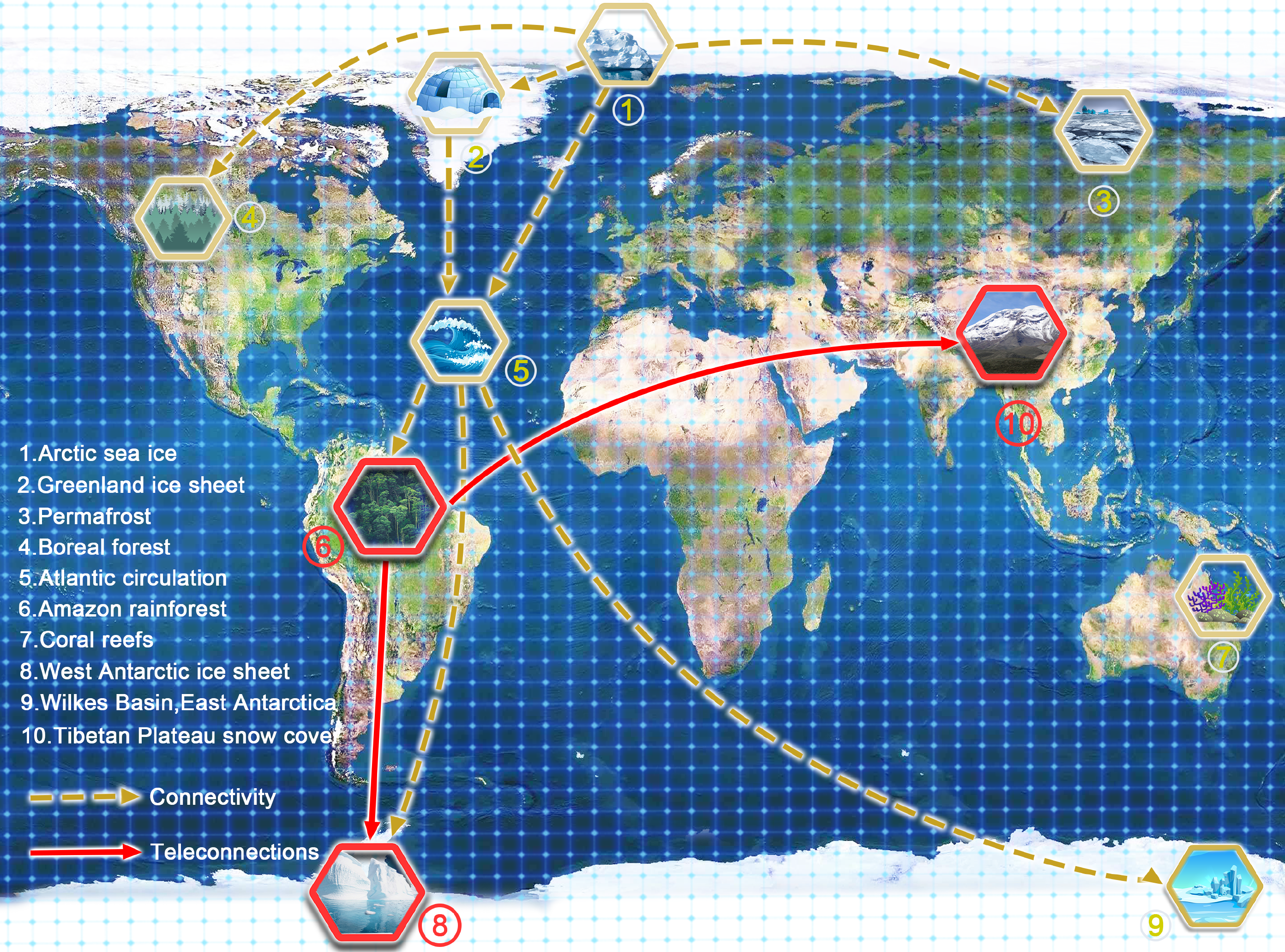}
\end{figure}
\noindent {\bf Fig. 1. Schematic view of the tipping elements of the Earth climate system, their connectivity and teleconnections.} The numbered symbols show the potential tipping elements in the Earth system. The dash yellow lines show the possible connections between these tipping elements, and the solid red lines show teleconnection uncovered in this article. The arrows show the direction of the influence.

\clearpage
\begin{figure}
\centering
\includegraphics[width=1\textwidth ]{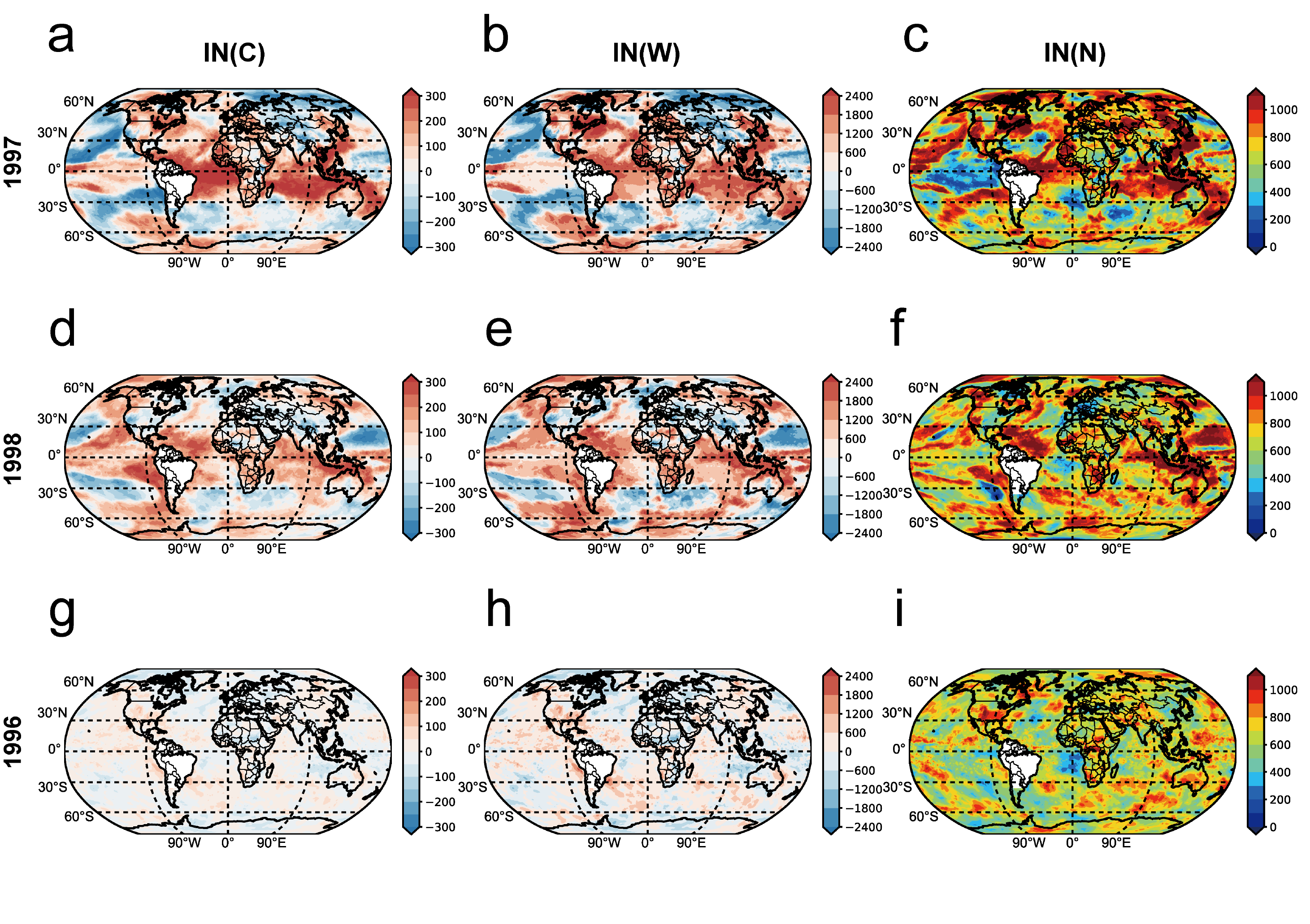}
\end{figure}
\noindent {\bf Fig. 2. The different influence modes of the ARA for ENSO years and a normal year.} $IN(C)$,   $IN(W)$ and $IN(N)$ show a more localized and higher intensity pattern in the \el~ ({\bf a-c}) and  \lanina~  ({\bf d-f}) years than in the normal year ({\bf g-i}).

\clearpage
\begin{figure}
\centering
\includegraphics[width=0.9\textwidth ]{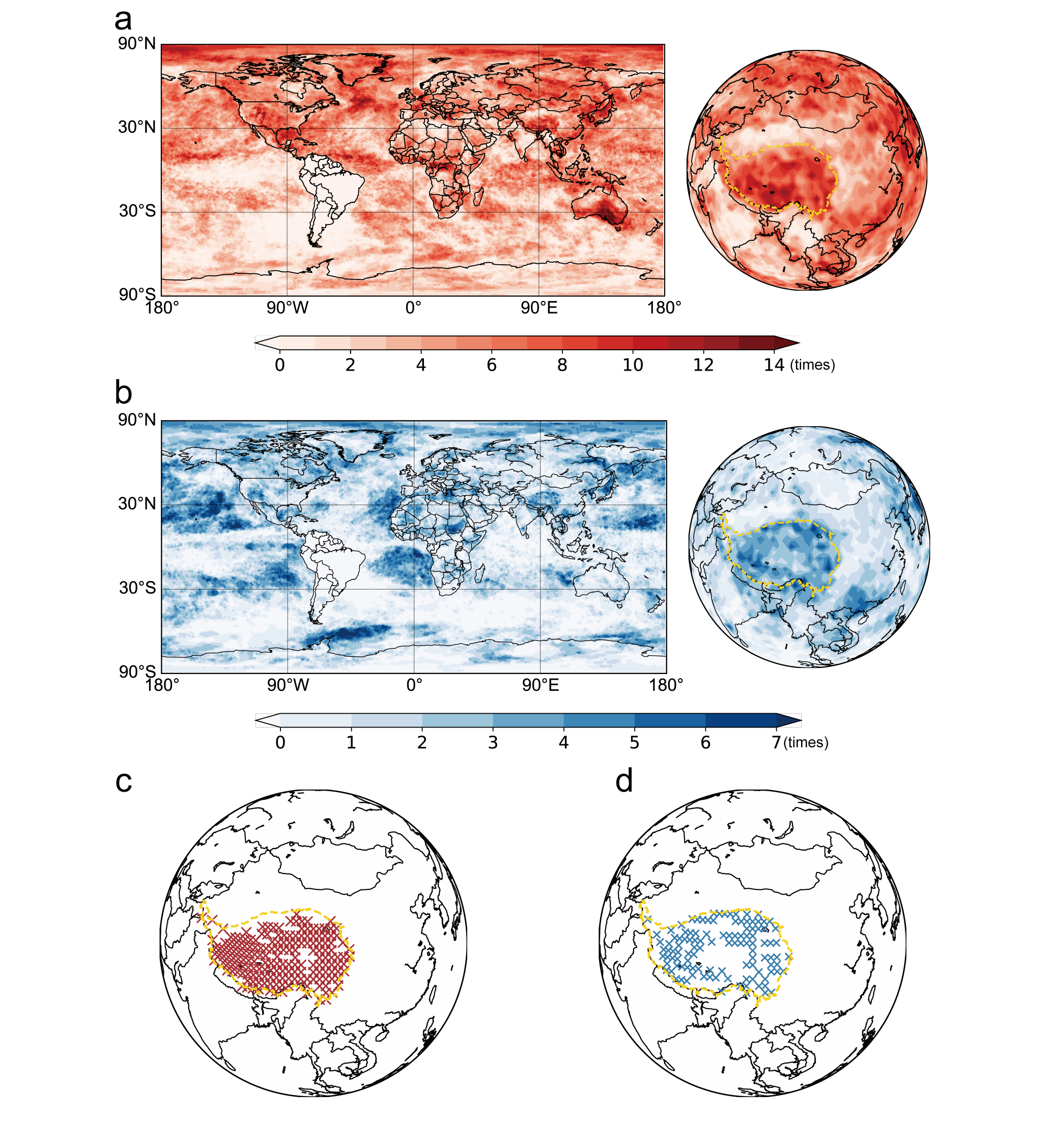}
\end{figure}
\noindent {\bf Fig. 3. Stable negative teleconnection between the Amazon area and the Tibetan Plateau.} {\bf a, b}, The spatial distribution of $F(N)$ and $F(C)$, depicting the areas influenced by ARA in past 40 years (1979-2018). The nodes within TP show high intensity, and the spatial pattern is perfectly characterised by the TP's cartographic boundary (the dashed orange line). {\bf c, d, } The crosses depict the nodes passing the hypothesis test. Here the red color indicates $F(N)$ and the blue color stands for $F(C)$.

\clearpage
\begin{figure}
\centering
\includegraphics[width=0.72\textwidth ]{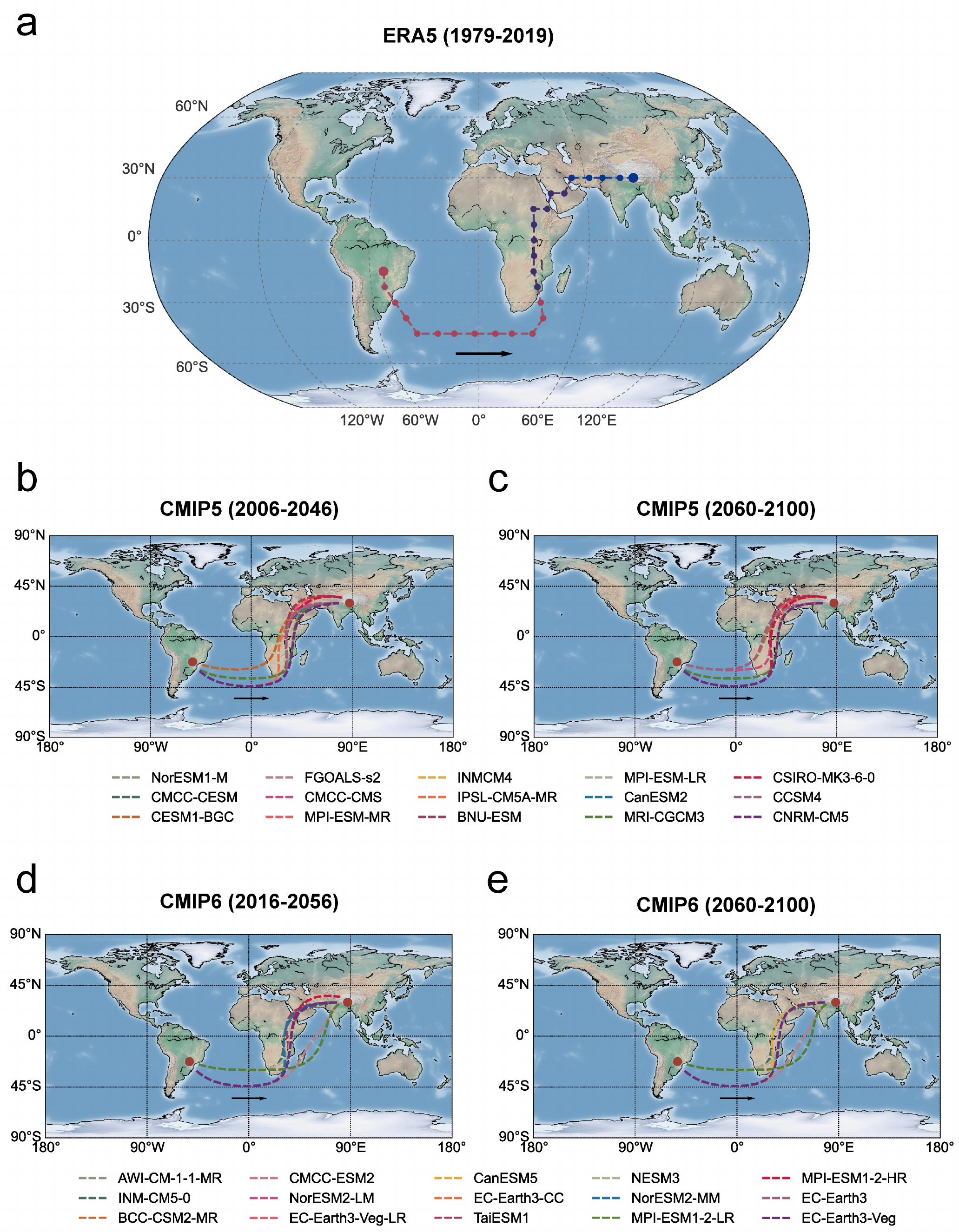}
\end{figure}
\noindent {\bf Fig. 4. The propagation pathway of the teleconnection between the Amazon area and the Tibetan Plateau.} {\bf a}, The large red and blue dots depict the starting and ending nodes, and the small dots are the network nodes passed along the path. The black arrow indicates the propagation direction. The potential meteorological interpretation of this path is described by three parts, corresponding to the dash lines with different colours. {\bf b-e}, The robust propagation pathway under global warming conditions. By comparing the pathway in the first ({\bf b, d}) and the last ({\bf c, e}) 41 years for this century in CMIP5 and CMIP6 datasets, we find that the overall pattern is quite stable across most models.

\clearpage
\begin{figure}
\centering
\includegraphics[width=0.9\textwidth ]{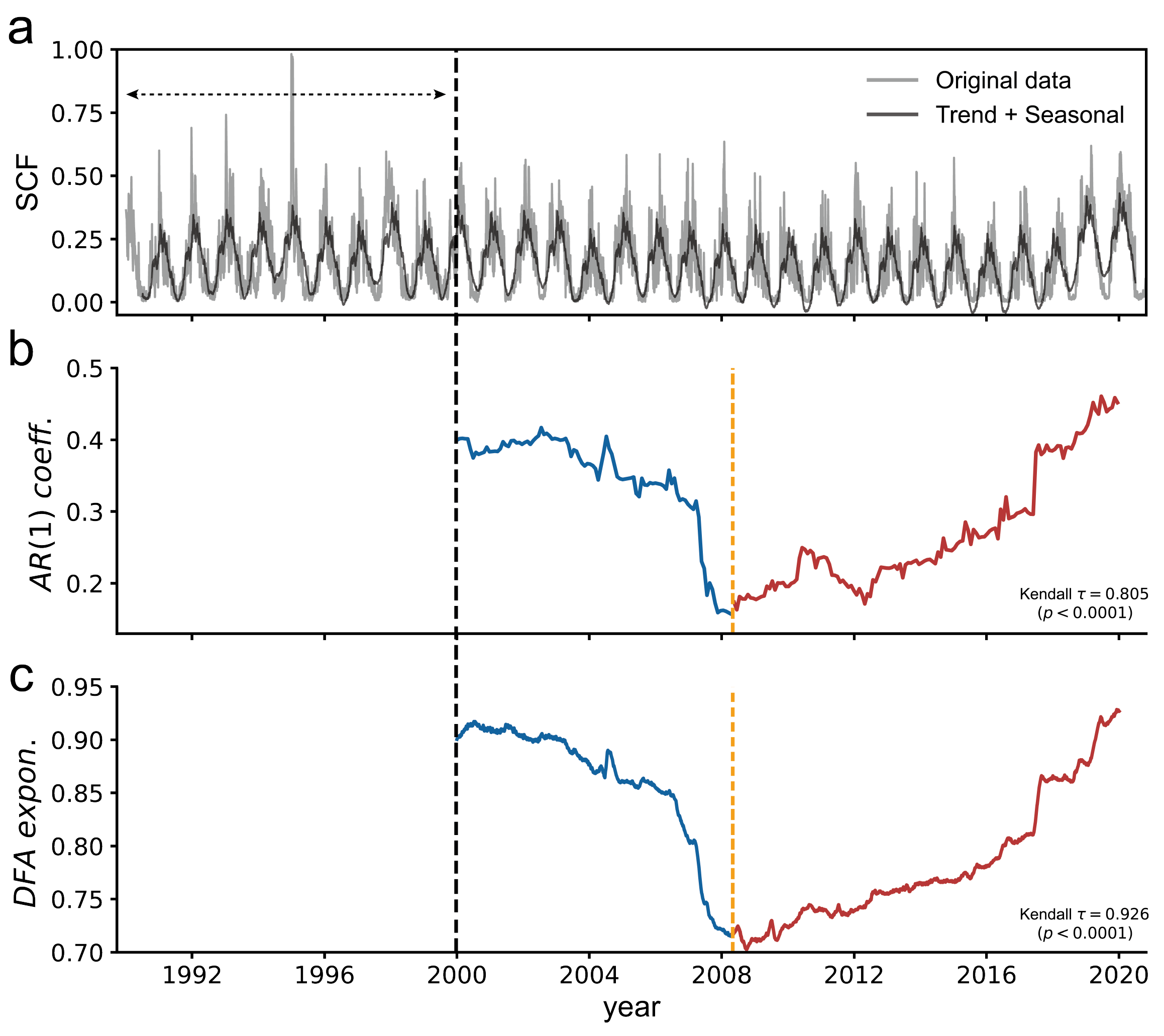}
\end{figure}
\noindent {\bf Fig. 5. The early warning indicators in Tibetan Plateau snow cover data show critical slowing down since 2008.} {\bf a}, The time series of the snow cover fraction (SCF) over the whole TP. Grey line shows the original data, black line shows the trend component and seasonal cycle component from the STL decomposed. {\bf b}, The time series of lag-1 autocorrelation (AR(1)) coefficient. {\bf c}, The time series of the detrended fluctuation analysis (DFA) exponent. AR(1) coefficient and DFA exponent are calculated using a sliding window with length 10 years. The black horizontal arrow represents the length of the moving window, and the vertical black dashed line indicates the time stamp from which the early warning measures are calculated. The line color depicts the different tendency of the time series.  The Kendall $\tau$ values of the AR(1) coefficient and DFA exponent since 2008 are 0.805 (p$<$0.0001) and 0.926 (p$<$0.0001), respectively. 
\clearpage

\section{Methods}
\subsection{Data}
Our climate network is based on the global hourly near-surface (1000 hPa) air temperature data from the ERA5 reanalysis dataset \cite{Hersbach2020} produced by the European Center for Medium-Range Weather Forecast (ECMWF). The reason we focus on the surface-temperature field is that it is the most commonly used for global-warming related discussions. The original resolution of the ERA5 dataset is $0.25^o\times 0.25^o$. We then reduce the spatial and temporal resolution by transforming the dataset to a resolution of $1^o\times 1^o$ and select the temperature at $00:00$ as the daily temperature value, resulting in 365 measurements for each of the resulting 360 × 181 = 65160 nodes for every year (in leap-years we exclude February 29th, thus all years have the same length). The dataset spans the time period from January 1979 to December 2019. In order to avoid the strong effect of seasonality, we subtract the calendar day’s mean from each nodes’ time series. The analysis of influence patterns of the ARA is based on a sequence of networks, each constructed from a time series that spans one year (the data in 2019 has been used for the time lag of the correlation calculation, so the networks cover 40 years, from 1979 to 2018). \\

\noindent To test whether the optimal pathway is independent of the specific dataset, we also employ surface air temperature reanalysis data ($0.25^o \times 0.25^o$) from ERA5, 1000 hPa air temperature reanalysis data ($2.5^o \times 2.5^o$) from NCEP/NCAR and surface temperature reanalysis data ($2.5^o \times 2.5^o$) from NCEP/NCAR.\\

\noindent We use a large set of climate models simulations from CMIP6 and CMIP5 to test the robustness of the teleconnection pathway.  Because of the different model resolutions, we apply an bilinear interpolation method to obtain new datasets with the same resolution of $2.5^o \times 2.5^o$. The outputs from CMIP5 were forced by Representative Concentration Pathway 8.5 (RCP8.5), covering a period from 2006 to 2100. The outputs from CMIP6 were forced by Shared Socioeconomic Pathway 5-8.5 (SSP 5-8.5), covering a period from 2016 to 2100. The variables of $\textit{ta}$ (air temperature) and $\textit{tas}$ (near-surface atmospheric temperature) are used here. The detailed information of these datasets can be found in Table. S1, S2.
\subsection{Climate network construction}
The nodes are divided into two subsets. One subset includes the nodes within the ARA (1374 nodes) and the other the nodes outside the ARA (63786 nodes). The links are constructed from the cross-correlation between two nodes from the different subsets. The cross-correlation values between the two time series of 365 days is defined by,
\begin{equation}
C_{ij}^y(\sigma) = \frac{\langle T_i(d)T_j(d+\sigma)\rangle-\langle T_i(d)\rangle\langle T_j(d+ \sigma)\rangle}{\sqrt{\langle(T_i(d)-\langle T_i(d)\rangle)^2\rangle}\cdot\sqrt{\langle(T_j(d+\sigma)-\langle T_j(d+\sigma)\rangle)^2\rangle}}
\end{equation}
Where $\sigma\in[0, \sigma_{max}]$ is the time lag, with $\sigma_{max} = $ 200 days, $y$ represents the starting year of this time series, and $C_{i,j}^y (-\sigma)\equiv C_{j,i}^y (\sigma)$. Therefore, we can achieve $2\sigma_{max}+1$ different cross-correlation values for every two nodes in one year. We then identify the maximum absolute value of this cross-correlation function and denote the corresponding time lag of this value as $[\sigma_0 ]_{i,j}^y$. The direction of each link is decided by the sign of $\sigma_0$. When the time lag is positive ($[\sigma_0 ]_{i,j}^y>0$), the direction of the link is from $i$ to $j$; when the time lag is negative, however, the direction is from $j$ to $i$. The link weights are determined by $C_{i,j}^y(\sigma_0)$, and we can also define the strength of the link $W_{i,j}^y$ as:
\begin{equation}
    W_{i,j}^y = \frac{C_{i,j}^y(\sigma_0)-mean(C_{i,j}^y(\sigma))}{std(C_{i,j}^y(\sigma))}.
\end{equation}
Where `mean' and `std' are the mean and SD of the cross-correlation function, respectively. We construct networks based on both $C_{i,j}^y(\sigma_0)$ and $W_{i,j}^y$. All results are indeed consistent with each other.\\

\noindent The in and out degree of each node can be calculated by $I_i^y=\sum_jA_{j,i}^y$ , $O_i^y=\sum_jA_{i,j}^y$, respectively. $A_{i,j}^y$ is the adjacency matrix of this network, and it is defined as:
\begin{equation}
    A_{i,j}^y = (1-\delta_{i,j})H([\sigma_0]_{i,j}^y).
\end{equation}
Where $H(x)$ is the Heaviside step function ($H(x\geq0)=1$ and $H(x<0)=0$). Furthermore, to describe the impact of the ARA on the global, we define the total in-degree of the node $j$ outside the ARA as the number of its in-links, the in-weights as the sum of the weights of its in-links, and the in-link strength as the sum of the strengths of its in-links:
\begin{equation}
    IN(N_j^y) = \sum_{i\in ARA}A_{i,j}^y,
\end{equation}
\begin{equation}
    IN(C_j^y) = \sum_{i\in ARA}A_{i,j}^yC_{i,j}^y(\sigma_0),
\end{equation}
\begin{equation}
    IN(W_j^y) = \sum_{i\in ARA}A_{i,j}^yW_{i,j}^y.
\end{equation}
The spatial distributions of $IN(C_j^y)$ and $IN(W_j^y)$ show the influence pattern of the ARA on the globe for a regarded year. Larger (smaller) positive (negative) values reflect stronger (weaker) warming (cooling) due to the impact of the ARA. In the same way, we can define out-degree, out-weights and out-strength to describe the impact of the outside world for the nodes in ARA:
\begin{equation}
    OUT(N_i^y) = \sum_{j\notin ARA}A_{j,i}^y,
\end{equation}
\begin{equation}
    OUT(C_j^y) = \sum_{j\notin ARA}A_{j,i}^yC_{j,i}^y(\sigma_0),
\end{equation}
\begin{equation}
    OUT(W_j^y) = \sum_{j\notin ARA}A_{j,i}^yW_{j,i}^y.
\end{equation}

\subsection{Filtering nodes from the network}
To extract the stable influenced region in the past 40 years, we removed insignificant nodes from the network. Since ENSO causes variation in the influence intensity, the links in a normal year are commonly weaker than that in an ENSO year. For this reason, a fixed threshold will remove most of the nodes in normal years. Therefore, we set an annual changing threshold, determined by the top 10\% of significant nodes in the current year. Besides, the nodes near the ARA always have a high positive link-weight, which will offer us trivial results after filtering. To avoid this, we focus on finding the regions with stable negative teleconnection to the ARA.\\

\noindent According to the sign of $IN(C_j^y)$, we divide the nodes outside the ARA into two subgroups: One subset includes the nodes with positive $IN(C_j^y)$, assigned as ${N_+^y}$; one subset includes the nodes with negative $IN(C_j^y)$, assigned as ${N_-^y}$. In order to find the most significant negative influenced nodes, we set a threshold to select:
\begin{equation}
    T_{j,-}^y = \mathcal{H}(IN(C_j^y)-C_-^y),\ j\in\{N_-^y\}.
\end{equation}
Where $\mathcal{H}$ is the Heaviside function, $C_-^y$ is determined by the value of $IN(C_j^y)$ in the top 10\% negative strength, i.e., corresponding to a significance of above 90\% confidence level. $T_{j,-}^y=1$ means the connection between node $j$ and ARA is stronger than of 90\% of the nodes in this year. Therefore, we can count the number of years for the node $j$ with top 10\% negative in-weight during the past 40 years:
\begin{equation}
    F(C_j) = \sum_{y}T_{j,-}^y.
\end{equation}
Similarily, $F(N_j)$ and $F(W_j)$ can be achieved from $IN(N_j^y)$ and $IN(W_j^y)$. The distribution of $F(N_j)$, $F(C_j)$ and $F(W_j)$ reflect the spatial distribution for the nodes that suffered persistent impact from the ARA.

\subsection{Optimal path finding}
We perform the shortest path method of complex networks to identify the optimal paths in our climate networks.
We select 726 nodes from the dataset to construct cross-correlation climate networks, and thus all nodes can approximately equally cover the globe. The distance between neighboring nodes is around 830km, and the distribution of the nodes can be found in Fig. S16. $1/|W_{i,j}^y|$ is defined as the cost value for each link to make sure the optimal path will prefer to pass the link with high significance \cite{zhou_teleconnection_2015}. The Dijkstra algorithm \cite{dijkstra_note_1959} was used to determine the directed optimal path between nodes $i$ and $j$ with the following limitations: (i) the distance for every step is shorter than 1000km; (ii) link time delay $\sigma_0 \ge 0$. The first limitation is used to identify the significant long-distance connections and the second limitation ensures that all steps have the same directions.

\subsection{Critical slowing down analysis}
Our critical slowing down analysis for the Tibet Plateau(TP) is based on a long-term Advanced Very High Resolution Radiometer (AVHRR) snow cover extent (SCE) dataset \cite{essd-13-4711-2021} from the Northwest Institute of Eco-Environment and Resources (NIEER), CAS. The dataset has a spatial resolution of 5km and a daily temporal resolution. Here we consider the data of the period from 1990 to 2020. To focus on changes in the snow cover of the whole TP, we use the time series of snow cover fraction (SCF) to measure the critical slowing down indicators. The SCF of the whole TP is calculated by:
\begin{equation}
r_{tp}(t) = \frac{n_{sc}(t)}{n_0},
\end{equation}
where $n_0 = 98549$ and $n_{sc}(t)$ are the number of grid boxes and the number of snow covered grid boxes in the Tibet Plateau, respectively.\\

\noindent To avoid false alarms, we use the Seasonal and Trend decomposition based on Loess (STL) method to filter out long-term trends and achieve stationarity \cite{lenton_early_2011,Dakos2008}. An STL function from Python has been used in this research to split $r_{tp}(t)$ into three parts, an overall trend, a repeating annual cycle, and a remaining residual. The remaining residual is  used in our analysis.
\subsubsection{The lag-1 autocorrelation (AR(1))}

The lag-1 autocorrelation (AR(1)) is a robust indicator for providing an EWS for an impending bifurcation-induced transitions and has been widely used~\cite{boulton_pronounced_2022,scheffer_anticipating_2012}. We measure our AR(1) coefficients based on the residual component of the decomposed snow cover daily time series.
Thirty-days average is applied to decrease the noise fluctuations. The time series of the AR(1) coefficient is obtained by a sliding time window with length 10 years.

\subsubsection{Detrended fluctuation analysis (DFA)}
The detrended fluctuation analysis (DFA) is another widely used tool for detecting the increase in memory caused by the critical slowing down \cite{Livina2007}, which can provide a useful cross-check of AR(1) \cite{lenton_early_2011}. For a time series with long-range temporal correlations, its fluctuation function, $F(n)$, can be characterized by a scaling exponent \cite{Peng1995}
\begin{equation}
    F(n) \sim n^{\alpha}.
\end{equation}
where $n$ is the window length, $\alpha$ is the DFA scaling exponent. Here, $F(n) = \sqrt{\frac{1}{n}\sum_{t=1}^n(X_t-Y_t^z)^2}$, where $X_t = \sum_{i=1}^t(x_i-\langle x \rangle)$ is the cumulative sum of the time series, and $Y_t^z$ is the fitted polynomial function, $z$ stands for the polynomial order (here we chose $z=2$). $F(n)$ is obtained by dividing the time series into $[L/n]$ non-overlaping time intervals of length n. The DFA exponent $\alpha$ is calculated as the slope of the linear fit to the log–log graph of $F(n)$ vs. $n$, for $10 \le n \le 1000$. The time series of DFA exponent $\alpha$ is obtained by a sliding time window with length 10 years.\\

\noindent We calculate the temporal trends of AR(1) coefficient and DFA exponent $\alpha$ by estimating the non-parametric Kendall rank correlation ($\tau$). Kendall $\tau$ is a statistical tool to measure the association between the variable and time. $\tau=1$ or $-1$ imply that the time series is always increasing or decreasing, $\tau=0$ means no overall trend. To test the robustness of the AR(1) and DFA analysis, we also vary the length of the sliding time window. The result with an alternatively 8-years time window is shown in Fig. S17, and we still see the same increase in AR(1) coefficient and DFA exponent. 

%\clearpage

{\section*{Data availability}
The ERA5 reanalysis data used here are publicly available at \url{https://cds.climate.copernicus.eu/cdsapp#!/dataset/reanalysis-era5-single-levels}. The NCEP/NCAR reanalysis data are publicly available at \url{https://psl.noaa.gov/data/gridded/data.ncep.reanalysis.html}. The CMIP5 data are publicly available at \url{https://esgf-node.llnl.gov/projects/cmip5/}. The CMIP6 data are publicly available at \url{https://esgf-node.llnl.gov/projects/cmip6/}. All other data that support the plots within this paper and other findings of this study are available from the corresponding author upon reasonable request.}
{\section*{Code availability}
The C++ and Python codes used for the analysis is available on GitHub (\url{https://github.com/fanjingfang/Tipping}).}

\section*{Acknowledgements}
The authors wish to thank Y. Ashkenazy for his helpful suggestions. This research is supported by grants from the National Natural Science Foundation of China (Grants No. 12275020, 12135003) and the Ministry of Science and Technology of China (Grants No. 2019QZKK0906).

{\section*{Author Contributions}
J.M, J.L, J.F, S.Y, D.C, J.K, X.C, S.H, H.J.S, designed the research, conceived the study, carried out the analysis, T.L and J.F performed the numerical calculations,
T.L, D.C, L.Y, Z.W, J.M, J.L, J.F, S.Y, D.C, J.K, X.C, S.H, H.J.S discussed results, and contributed to writing the manuscript.  J.F. led the writing of the manuscript.}

{\section*{Additional information}
Supplementary Information is available in the online version of the paper.}

\section*{Competing interests}
The authors declare no competing  interests.

\bibliographystyle{naturemag}
\bibliography{MyLibrary}

\begin{thebibliography}{10}
\expandafter\ifx\csname url\endcsname\relax
  \def\url#1{\texttt{#1}}\fi
\expandafter\ifx\csname urlprefix\endcsname\relax\def\urlprefix{URL }\fi
\providecommand{\bibinfo}[2]{#2}
\providecommand{\eprint}[2][]{\url{#2}}

\bibitem{lenton_tipping_2008}
\bibinfo{author}{Lenton, T.~M.} \emph{et~al.}
\newblock \bibinfo{title}{Tipping elements in the {Earth}'s climate system}.
\newblock \emph{\bibinfo{journal}{Proceedings of the National Academy of
  Sciences}} \textbf{\bibinfo{volume}{105}}, \bibinfo{pages}{1786--1793}
  (\bibinfo{year}{2008}).

\bibitem{lenton_climate_2019}
\bibinfo{author}{Lenton, T.~M.} \emph{et~al.}
\newblock \bibinfo{title}{Climate tipping points — too risky to bet against}.
\newblock \emph{\bibinfo{journal}{Nature}} \textbf{\bibinfo{volume}{575}},
  \bibinfo{pages}{592--595} (\bibinfo{year}{2019}).

\bibitem{lovejoy2019amazon}
\bibinfo{author}{Lovejoy, T.~E.} \& \bibinfo{author}{Nobre, C.}
\newblock \bibinfo{title}{Amazon tipping point: Last chance for action}.
\newblock \emph{\bibinfo{journal}{Science Advances}}
  \textbf{\bibinfo{volume}{5}}, \bibinfo{pages}{eaba2949}
  (\bibinfo{year}{2019}).

\bibitem{taubert_global_2018}
\bibinfo{author}{Taubert, F.} \emph{et~al.}
\newblock \bibinfo{title}{Global patterns of tropical forest fragmentation}.
\newblock \emph{\bibinfo{journal}{Nature}}  (\bibinfo{year}{2018}).

\bibitem{boers_observation-based_2021}
\bibinfo{author}{Boers, N.}
\newblock \bibinfo{title}{Observation-based early-warning signals for a
  collapse of the {Atlantic} {Meridional} {Overturning} {Circulation}}.
\newblock \emph{\bibinfo{journal}{Nature Climate Change}}
  \textbf{\bibinfo{volume}{11}}, \bibinfo{pages}{680--688}
  (\bibinfo{year}{2021}).

\bibitem{garbe_hysteresis_2020}
\bibinfo{author}{Garbe, J.}, \bibinfo{author}{Albrecht, T.},
  \bibinfo{author}{Levermann, A.}, \bibinfo{author}{Donges, J.~F.} \&
  \bibinfo{author}{Winkelmann, R.}
\newblock \bibinfo{title}{The hysteresis of the {Antarctic} {Ice} {Sheet}}.
\newblock \emph{\bibinfo{journal}{Nature}} \textbf{\bibinfo{volume}{585}},
  \bibinfo{pages}{538--544} (\bibinfo{year}{2020}).

\bibitem{meredith2019polar}
\bibinfo{author}{Meredith, M.} \emph{et~al.}
\newblock \bibinfo{title}{Polar {Regions}. {Chapter} 3, {IPCC} {Special}
  {Report} on the {Ocean} and {Cryosphere} in a {Changing} {Climate}}
  (\bibinfo{year}{2019}).

\bibitem{ghil_physics_2020}
\bibinfo{author}{Ghil, M.} \& \bibinfo{author}{Lucarini, V.}
\newblock \bibinfo{title}{The physics of climate variability and climate
  change}.
\newblock \emph{\bibinfo{journal}{Reviews of Modern Physics}}
  \textbf{\bibinfo{volume}{92}}, \bibinfo{pages}{035002}.

\bibitem{lenton_early_2011}
\bibinfo{author}{Lenton, T.~M.}
\newblock \bibinfo{title}{Early warning of climate tipping points}.
\newblock \emph{\bibinfo{journal}{Nature Climate Change}}
  \textbf{\bibinfo{volume}{1}}, \bibinfo{pages}{201--209}
  (\bibinfo{year}{2011}).

\bibitem{IPCC6}
\bibinfo{author}{Masson-Delmotte} \emph{et~al.}
\newblock \bibinfo{title}{{IPCC}, 2021: {Climate Change} 2021: The {Physical}
  {Science} {Basis}. {Contribution} of {Working} {Group} {I} to the {Sixth}
  {Assessment} {Report} of the {Intergovernmental} {Panel} on {Climate}
  {Change}}  (\bibinfo{year}{2021}).

\bibitem{Zhang2020}
\bibinfo{author}{Zhang, P.} \emph{et~al.}
\newblock \bibinfo{title}{{Abrupt} shift to hotter and drier climate over inner
  {East Asia} beyond the tipping point}.
\newblock \emph{\bibinfo{journal}{Science}} \textbf{\bibinfo{volume}{370}},
  \bibinfo{pages}{1095--1099} (\bibinfo{year}{2020}).

\bibitem{scheffer_anticipating_2012}
\bibinfo{author}{Scheffer, M.} \emph{et~al.}
\newblock \bibinfo{title}{Anticipating {Critical} {Transitions}}.
\newblock \emph{\bibinfo{journal}{Science}} \textbf{\bibinfo{volume}{338}},
  \bibinfo{pages}{344--348} (\bibinfo{year}{2012}).

\bibitem{Klose2021}
\bibinfo{author}{Klose, A.~K.}, \bibinfo{author}{Wunderling, N.},
  \bibinfo{author}{Winkelmann, R.} \& \bibinfo{author}{Donges, J.~F.}
\newblock \bibinfo{title}{What do we mean, `tipping cascade'?}
\newblock \emph{\bibinfo{journal}{Environmental Research Letters}}
  \textbf{\bibinfo{volume}{16}}, \bibinfo{pages}{125011}
  (\bibinfo{year}{2021}).

\bibitem{Brovkin2021}
\bibinfo{author}{Brovkin, V.} \emph{et~al.}
\newblock \bibinfo{title}{Past abrupt changes, tipping points and cascading
  impacts in the earth system}.
\newblock \emph{\bibinfo{journal}{Nature Geoscience}}
  \textbf{\bibinfo{volume}{14}}, \bibinfo{pages}{550--558}
  (\bibinfo{year}{2021}).

\bibitem{Steffen2018}
\bibinfo{author}{Steffen, W.} \emph{et~al.}
\newblock \bibinfo{title}{Trajectories of the earth system in the
  anthropocene}.
\newblock \emph{\bibinfo{journal}{Proceedings of the National Academy of
  Sciences}} \textbf{\bibinfo{volume}{115}}, \bibinfo{pages}{8252--8259}
  (\bibinfo{year}{2018}).

\bibitem{martin2021}
\bibinfo{author}{Martin, M.~A.} \emph{et~al.}
\newblock \bibinfo{title}{Ten new insights in climate science 2021: a horizon
  scan}.
\newblock \emph{\bibinfo{journal}{Global Sustainability}}
  \textbf{\bibinfo{volume}{4}}, \bibinfo{pages}{e25} (\bibinfo{year}{2021}).

\bibitem{gibson_primary_2011}
\bibinfo{author}{Gibson, L.} \emph{et~al.}
\newblock \bibinfo{title}{Primary forests are irreplaceable for sustaining
  tropical biodiversity}.
\newblock \emph{\bibinfo{journal}{Nature}} \textbf{\bibinfo{volume}{478}},
  \bibinfo{pages}{378--381} (\bibinfo{year}{2011}).

\bibitem{achard_determination_2014}
\bibinfo{author}{Achard, F.} \emph{et~al.}
\newblock \bibinfo{title}{Determination of tropical deforestation rates and
  related carbon losses from 1990 to 2010}.
\newblock \emph{\bibinfo{journal}{Global Change Biology}}
  \textbf{\bibinfo{volume}{20}}, \bibinfo{pages}{2540--2554}
  (\bibinfo{year}{2014}).

\bibitem{hansen_high-resolution_2013}
\bibinfo{author}{Hansen, M.~C.} \emph{et~al.}
\newblock \bibinfo{title}{High-{Resolution} {Global} {Maps} of 21st-{Century}
  {Forest} {Cover} {Change}}.
\newblock \emph{\bibinfo{journal}{Science}} \textbf{\bibinfo{volume}{342}},
  \bibinfo{pages}{850--853}.

\bibitem{gatti_amazonia_2021}
\bibinfo{author}{Gatti, L.~V.} \emph{et~al.}
\newblock \bibinfo{title}{Amazonia as a carbon source linked to deforestation
  and climate change}.
\newblock \emph{\bibinfo{journal}{Nature}} \textbf{\bibinfo{volume}{595}},
  \bibinfo{pages}{388--393} (\bibinfo{year}{2021}).

\bibitem{boulton_pronounced_2022}
\bibinfo{author}{Boulton, C.~A.}, \bibinfo{author}{Lenton, T.~M.} \&
  \bibinfo{author}{Boers, N.}
\newblock \bibinfo{title}{Pronounced loss of {Amazon} rainforest resilience
  since the early 2000s}.
\newblock \emph{\bibinfo{journal}{Nature Climate Change}}
  \textbf{\bibinfo{volume}{12}}, \bibinfo{pages}{1--8} (\bibinfo{year}{2022}).

\bibitem{nobre_land-use_2016}
\bibinfo{author}{Nobre, C.~A.} \emph{et~al.}
\newblock \bibinfo{title}{Land-use and climate change risks in the {Amazon} and
  the need of a novel sustainable development paradigm}.
\newblock \emph{\bibinfo{journal}{Proceedings of the National Academy of
  Sciences}} \textbf{\bibinfo{volume}{113}}, \bibinfo{pages}{10759--10768}
  (\bibinfo{year}{2016}).

\bibitem{newman2010networks}
\bibinfo{author}{Newman, M. E.~J.}
\newblock \emph{\bibinfo{title}{Networks: {An} {Introduction}}}
  (\bibinfo{publisher}{Oxford University Press}, \bibinfo{address}{Oxford, UK},
  \bibinfo{year}{2010}).

\bibitem{tsonis_architecture_2004}
\bibinfo{author}{Tsonis, A.~A.} \& \bibinfo{author}{Roebber, P.~J.}
\newblock \bibinfo{title}{The architecture of the climate network}.
\newblock \emph{\bibinfo{journal}{Physica A: Statistical Mechanics and its
  Applications}} \textbf{\bibinfo{volume}{333}}, \bibinfo{pages}{497--504}
  (\bibinfo{year}{2004}).

\bibitem{ludescher_improved_2013}
\bibinfo{author}{Ludescher, J.} \emph{et~al.}
\newblock \bibinfo{title}{Improved {El} {Ni\~no} forecasting by cooperativity
  detection}.
\newblock \emph{\bibinfo{journal}{Proceedings of the National Academy of
  Sciences}} \textbf{\bibinfo{volume}{110}}, \bibinfo{pages}{11742--11745}
  (\bibinfo{year}{2013}).

\bibitem{boers_complex_2019}
\bibinfo{author}{Boers, N.} \emph{et~al.}
\newblock \bibinfo{title}{Complex networks reveal global pattern of
  extreme-rainfall teleconnections}.
\newblock \emph{\bibinfo{journal}{Nature}} \textbf{\bibinfo{volume}{566}},
  \bibinfo{pages}{373--377} (\bibinfo{year}{2019}).

\bibitem{NetworkMonsoon}
\bibinfo{author}{Fan, J.} \emph{et~al.}
\newblock \bibinfo{title}{Network-{Based} {Approach} and {Climate Change}
  {Benefits} for {Forecasting} the {Amount } of {Indian} {Monsoon} {Rainfall}}.
\newblock \emph{\bibinfo{journal}{Journal of Climate}}
  \textbf{\bibinfo{volume}{35}}, \bibinfo{pages}{1 -- 39}
  (\bibinfo{year}{2021}).

\bibitem{mheen_interaction_2013}
\bibinfo{author}{Mheen, M. v.~d.} \emph{et~al.}
\newblock \bibinfo{title}{Interaction network based early warning indicators
  for the {Atlantic} {MOC} collapse}.
\newblock \emph{\bibinfo{journal}{Geophysical Research Letters}}
  \textbf{\bibinfo{volume}{40}}, \bibinfo{pages}{2714--2719}
  (\bibinfo{year}{2013}).

\bibitem{feng_are_2014}
\bibinfo{author}{Feng, Q.~Y.} \& \bibinfo{author}{Dijkstra, H.}
\newblock \bibinfo{title}{Are {North} {Atlantic} multidecadal {SST} anomalies
  westward propagating?}
\newblock \emph{\bibinfo{journal}{Geophysical Research Letters}}
  \textbf{\bibinfo{volume}{41}}, \bibinfo{pages}{541--546}
  (\bibinfo{year}{2014}).

\bibitem{fan_statistical_2021}
\bibinfo{author}{Fan, J.} \emph{et~al.}
\newblock \bibinfo{title}{Statistical physics approaches to the complex {Earth}
  system}.
\newblock \emph{\bibinfo{journal}{Physics Reports}}
  \textbf{\bibinfo{volume}{896}}, \bibinfo{pages}{1--84}
  (\bibinfo{year}{2021}).

\bibitem{yao_recent_2019}
\bibinfo{author}{Yao, T.} \emph{et~al.}
\newblock \bibinfo{title}{Recent {Third} {Pole}’s {Rapid} {Warming}
  {Accompanies} {Cryospheric} {Melt} and {Water} {Cycle} {Intensification} and
  {Interactions} between {Monsoon} and {Environment}: {Multidisciplinary}
  {Approach} with {Observations}, {Modeling}, and {Analysis}}.
\newblock \emph{\bibinfo{journal}{Bulletin of the American Meteorological
  Society}} \textbf{\bibinfo{volume}{100}}, \bibinfo{pages}{423--444}
  (\bibinfo{year}{2019}).

\bibitem{cai_climate_2020}
\bibinfo{author}{Cai, W.} \emph{et~al.}
\newblock \bibinfo{title}{Climate impacts of the {El} {Niño}–{Southern}
  {Oscillation} on {South} {America}}.
\newblock \emph{\bibinfo{journal}{Nature Reviews Earth \& Environment}}
  \textbf{\bibinfo{volume}{1}}, \bibinfo{pages}{215--231}
  (\bibinfo{year}{2020}).

\bibitem{Zhengyu2007}
\bibinfo{author}{Liu, Z.} \& \bibinfo{author}{Alexander, M.}
\newblock \bibinfo{title}{Atmospheric bridge, oceanic tunnel, and global
  climatic teleconnections}.
\newblock \emph{\bibinfo{journal}{Reviews of Geophysics}}
  \textbf{\bibinfo{volume}{45}} (\bibinfo{year}{2007}).

\bibitem{zhou_teleconnection_2015}
\bibinfo{author}{Zhou, D.}, \bibinfo{author}{Gozolchiani, A.},
  \bibinfo{author}{Ashkenazy, Y.} \& \bibinfo{author}{Havlin, S.}
\newblock \bibinfo{title}{Teleconnection {Paths} via {Climate} {Network}
  {Direct} {Link} {Detection}}.
\newblock \emph{\bibinfo{journal}{Physical Review Letters}}
  \textbf{\bibinfo{volume}{115}}, \bibinfo{pages}{268501}
  (\bibinfo{year}{2015}).

\bibitem{leduc1984connaitre}
\bibinfo{author}{Leduc, R.} \& \bibinfo{author}{Gervais, R.}
\newblock \emph{\bibinfo{title}{Conna{\^\i}tre la m{\'e}t{\'e}orologie}}
  (\bibinfo{publisher}{PUQ}, \bibinfo{year}{1984}).

\bibitem{nicholson_itcz_2018}
\bibinfo{author}{Nicholson, S.~E.}
\newblock \bibinfo{title}{The {ITCZ} and the {Seasonal} {Cycle} over
  {Equatorial} {Africa}}.
\newblock \emph{\bibinfo{journal}{Bulletin of the American Meteorological
  Society}} \textbf{\bibinfo{volume}{99}}, \bibinfo{pages}{337--348}
  (\bibinfo{year}{2018}).

\bibitem{kong_interaction_2020}
\bibinfo{author}{Kong, W.} \& \bibinfo{author}{Chiang, J. C.~H.}
\newblock \bibinfo{title}{Interaction of the {Westerlies} with the {Tibetan}
  {Plateau} in {Determining} the {Mei}-{Yu} {Termination}}.
\newblock \emph{\bibinfo{journal}{Journal of Climate}}
  \textbf{\bibinfo{volume}{33}}, \bibinfo{pages}{339--363}
  (\bibinfo{year}{2020}).

\bibitem{Feldmann14191}
\bibinfo{author}{Feldmann, J.} \& \bibinfo{author}{Levermann, A.}
\newblock \bibinfo{title}{{Collapse} of the {West Antarctic Ice Sheet} after
  local destabilization of the {Amundsen Basin}}.
\newblock \emph{\bibinfo{journal}{Proceedings of the National Academy of
  Sciences}} \textbf{\bibinfo{volume}{112}}, \bibinfo{pages}{14191--14196}
  (\bibinfo{year}{2015}).

\bibitem{pattyn_uncertain_2020}
\bibinfo{author}{Pattyn, F.} \& \bibinfo{author}{Morlighem, M.}
\newblock \bibinfo{title}{The uncertain future of the {Antarctic} {Ice}
  {Sheet}}.
\newblock \emph{\bibinfo{journal}{Science}} \textbf{\bibinfo{volume}{367}},
  \bibinfo{pages}{1331--1335} (\bibinfo{year}{2020}).

\bibitem{McConnell2007}
\bibinfo{author}{McConnell, J.~R.}, \bibinfo{author}{Aristarain, A.~J.},
  \bibinfo{author}{Banta, J.~R.}, \bibinfo{author}{Edwards, P.~R.} \&
  \bibinfo{author}{Simoes, J.~C.}
\newblock \bibinfo{title}{20th-century doubling in dust archived in an
  {Antarctic Peninsula} ice core parallels climate change and desertification
  in {South America}}.
\newblock \emph{\bibinfo{journal}{Proceedings of the National Academy of
  Sciences}} \textbf{\bibinfo{volume}{104}}, \bibinfo{pages}{5743--5748}
  (\bibinfo{year}{2007}).

\bibitem{kim_assessment_2021}
\bibinfo{author}{Kim, G.} \emph{et~al.}
\newblock \bibinfo{title}{Assessment of {MME} methods for seasonal prediction
  using {WMO} {LC}‐{LRFMME} hindcast dataset}.
\newblock \emph{\bibinfo{journal}{International Journal of Climatology}}
  \textbf{\bibinfo{volume}{41}}, \bibinfo{pages}{E2462--E2481}
  (\bibinfo{year}{2021}).

\bibitem{you_warming_2021}
\bibinfo{author}{You, Q.} \emph{et~al.}
\newblock \bibinfo{title}{Warming amplification over the {Arctic} {Pole} and
  {Third} {Pole}: {Trends}, mechanisms and consequences}.
\newblock \emph{\bibinfo{journal}{Earth-Science Reviews}}
  \textbf{\bibinfo{volume}{217}}, \bibinfo{pages}{103625}
  (\bibinfo{year}{2021}).

\bibitem{you_tibetan_2020}
\bibinfo{author}{You, Q.} \emph{et~al.}
\newblock \bibinfo{title}{Tibetan {Plateau} amplification of climate extremes
  under global warming of 1.5 °{C}, 2 °{C} and 3 °{C}}.
\newblock \emph{\bibinfo{journal}{Global and Planetary Change}}
  \textbf{\bibinfo{volume}{192}}, \bibinfo{pages}{103261}
  (\bibinfo{year}{2020}).

\bibitem{ye1998role}
\bibinfo{author}{Ye, D.~Z.} \& \bibinfo{author}{Wu, G.~X.}
\newblock \bibinfo{title}{The role of the heat source of the {Tibetan Plateau}
  in the general circulation}.
\newblock \emph{\bibinfo{journal}{Meteorology and Atmospheric Physics}}
  \textbf{\bibinfo{volume}{67}}, \bibinfo{pages}{181--198}
  (\bibinfo{year}{1998}).

\bibitem{YANG1996}
\bibinfo{author}{Yang, S.}
\newblock \bibinfo{title}{{ENSO}–snow–monsoon associations and
  seasonal–interannual predictions}.
\newblock \emph{\bibinfo{journal}{International Journal of Climatology}}
  \textbf{\bibinfo{volume}{16}}, \bibinfo{pages}{125--134}
  (\bibinfo{year}{1996}).

\bibitem{Qin2006}
\bibinfo{author}{Dahe, Q.}, \bibinfo{author}{Shiyin, L.} \&
  \bibinfo{author}{Peiji, L.}
\newblock \bibinfo{title}{Snow cover distribution, variability, and response to
  climate change in western china}.
\newblock \emph{\bibinfo{journal}{Journal of Climate}}
  \textbf{\bibinfo{volume}{19}}, \bibinfo{pages}{1820 -- 1833}
  (\bibinfo{year}{2006}).

\bibitem{ditlevsen_tipping_2010}
\bibinfo{author}{Ditlevsen, P.~D.} \& \bibinfo{author}{Johnsen, S.~J.}
\newblock \bibinfo{title}{Tipping points: {Early} warning and wishful
  thinking}.
\newblock \emph{\bibinfo{journal}{Geophysical Research Letters}}
  \textbf{\bibinfo{volume}{37}}, \bibinfo{pages}{L19703}
  (\bibinfo{year}{2010}).

\bibitem{Peng1995}
\bibinfo{author}{Peng, C.}, \bibinfo{author}{Havlin, S.},
  \bibinfo{author}{Stanley, H.~E.} \& \bibinfo{author}{Goldberger, A.~L.}
\newblock \bibinfo{title}{Quantification of scaling exponents and crossover
  phenomena in nonstationary heartbeat time series}.
\newblock \emph{\bibinfo{journal}{Chaos: An Interdisciplinary Journal of
  Nonlinear Science}} \textbf{\bibinfo{volume}{5}}, \bibinfo{pages}{82--87}
  (\bibinfo{year}{1995}).

\bibitem{Livina2007}
\bibinfo{author}{Livina, V.~N.} \& \bibinfo{author}{Lenton, T.~M.}
\newblock \bibinfo{title}{A modified method for detecting incipient
  bifurcations in a dynamical system}.
\newblock \emph{\bibinfo{journal}{Geophysical Research Letters}}
  \textbf{\bibinfo{volume}{34}} (\bibinfo{year}{2007}).

\bibitem{Dakos2008}
\bibinfo{author}{Dakos, V.} \emph{et~al.}
\newblock \bibinfo{title}{Slowing down as an early warning signal for abrupt
  climate change}.
\newblock \emph{\bibinfo{journal}{Proceedings of the National Academy of
  Sciences}} \textbf{\bibinfo{volume}{105}}, \bibinfo{pages}{14308--14312}
  (\bibinfo{year}{2008}).

\bibitem{carleton_social_2016}
\bibinfo{author}{Carleton, T.~A.} \& \bibinfo{author}{Hsiang, S.~M.}
\newblock \bibinfo{title}{Social and economic impacts of climate}.
\newblock \emph{\bibinfo{journal}{Science}} \textbf{\bibinfo{volume}{353}},
  \bibinfo{pages}{aad9837} (\bibinfo{year}{2016}).

\bibitem{Hersbach2020}
\bibinfo{author}{Hersbach, H.} \emph{et~al.}
\newblock \bibinfo{title}{The {ERA5} global reanalysis}.
\newblock \emph{\bibinfo{journal}{Quarterly Journal of the Royal Meteorological
  Society}} \textbf{\bibinfo{volume}{146}}, \bibinfo{pages}{1999--2049}
  (\bibinfo{year}{2020}).

\bibitem{dijkstra_note_1959}
\bibinfo{author}{Dijkstra, E.~W.}
\newblock \bibinfo{title}{A note on two problems in connexion with graphs}.
\newblock \emph{\bibinfo{journal}{Numerische Mathematik}}
  \textbf{\bibinfo{volume}{1}}, \bibinfo{pages}{269--271}
  (\bibinfo{year}{1959}).

\bibitem{essd-13-4711-2021}
\bibinfo{author}{Hao, X.} \emph{et~al.}
\newblock \bibinfo{title}{The {NIEER AVHRR} snow cover extent product over
  {China} -- a long-term daily snow record for regional climate research}.
\newblock \emph{\bibinfo{journal}{Earth System Science Data}}
  \textbf{\bibinfo{volume}{13}}, \bibinfo{pages}{4711--4726}
  (\bibinfo{year}{2021}).

\end{thebibliography}
\end{document}